\documentclass[intlimits,twoside,a4paper]{article}

\usepackage{amsmath,amssymb}
\usepackage{graphicx}
\usepackage{hhline}

\usepackage[T2A]{fontenc}
\usepackage[cp1251]{inputenc}

\usepackage[eqsecnum]{cmpj2}

\issue{2017}{20}{1}{13901}
\doinumber{10.5488/CMP.20.13901}

\title[Voids in Cosmic Web]%
{Voids in the Cosmic Web as a probe of dark energy\thanks{On the occasion of the 60th anniversary of our true friend, colleague and prominent scientist Yurko Holovatch.}}
\author{B. Novosyadlyj, M. Tsizh}
\address{Astronomical Observatory, Ivan Franko National University of Lviv,
\\ 8~Kyryla i Methodia St., 79005 Lviv, Ukraine}

\authorcopyright{B. Novosyadlyj, M. Tsizh, 2017}
\date{Received January 17, 2017, in final form February 27, 2017}

\begin{document}

\maketitle

\begin{abstract}
The formation of large voids in the Cosmic Web from the initial adiabatic cosmological
perturbations of space-time metric, density and velocity of matter
is investigated in cosmological model with the dynamical dark energy accelerating
expansion of the Universe. It is shown that the negative density
perturbations with the initial radius of about 50~Mpc in comoving to the cosmological
background coordinates and the amplitude corresponding to the r.m.s.
temperature fluctuations of the cosmic microwave
background lead to the formation of voids with the density contrast up to $-$0.9,
maximal peculiar velocity about 400~km/s and the radius close to the initial one.
An important feature of voids formation from the analyzed initial amplitudes
and profiles is establishing the surrounding overdensity shell. We have shown
that the ratio of the peculiar velocity in units of the Hubble flow to the density contrast in the
central part of a void does not depend or weakly depends on the distance from
the center of the void. It is also shown that this ratio is sensitive to the values
of dark energy parameters and can be used to find them based on the
observational data on mass density and peculiar velocities of galaxies in the
voids.
\keywords cosmological perturbations, dark energy, large scale structure of the
Universe, voids, Cosmic Web
\pacs 95.36.+x, 98.80.-k
\end{abstract}
\vspace{-6mm}
\section{Introduction}
\vspace{-1mm}
The cosmological constant, which has been introduced into general relativity one
hundred years ago by Albert Einstein \cite{Einstein1917}, played an important
role in the development of the theory of structure and evolution of the
Universe. Late in the 20th century, two research teams, SuperNova Cosmology
Project and High-z SuperNova Search, performed the cosmological test
``magnitude--redshift'' for type Ia supernovae and obtained secure
observational evidence (at 3$\sigma$ confidence level) for non-zero positive value of the
cosmological
constant \cite{Perlmutter1999,Riess1998,Schmidt1998} and accelerated expansion
of the Universe as a consequence. Since the cosmological constant has no
consistent physical interpretation, other essences have been proposed to
explain
the accelerated expansion of the Universe. They have been
commonly referred to as the ``dark energy'' \cite{Huterer1999}. Among them the most elaborated ones
are the scalar field models of dark energy
\cite{Amendola2010,Wolschin2010,Ruiz2010,Novosyadlyj2013m}. In the ``world
ocean'' of different variants of such models one can distinguish a few ``main
streams'', such as quintessential, phantom, quintom and others. Accurate
observations
and new crucial tests are needed to discriminate between these candidates and
decide which one fills our Universe. In numerous papers of the last decade it
was
shown that large voids in a large-scale structure of the Universe are
excellent
laboratories for this purpose
\citep{Li09,Biswas10,Bos12,Li12,Clampitt13,Jennings13,Cai15,Dai15,Hamaus15,
Tsizh2016,Novosyadlyj2017}. Here, we show that the data on dynamical structure of
the voids, positions, number densities and peculiar velocities of the galaxies at
central parts of
the voids are sensitive to the parameters of dark energy.

\vspace{-3mm}
\section{Voids in Cosmic Web}
\vspace{-1mm}

The space distribution of luminous matter, which is collected in galaxies,
forms a large-scale structure of the Universe. Its elements --- clusters,
superclusters, walls, sheets, filaments, nodes and voids~--- form a complex
geometrical and topological system which is sometimes dubbed as the Cosmic Web
\cite{Bond96}. It emerged from the quantum fluctuations of space-time metric
generated in the early universe and stretched to the cosmic scale by inflation. Its
formation in the expanding Universe
is a long history of action of the forces of self-gravitation of luminous matter,
gravity of dark matter and ``anti-gravity'' of dark energy on large scales. The
complexity of the structure
requires sophisticated methods of analysis. Methods of complex systems
analysis which have been recently developed \cite{Albert2002, Holovach06, Holovach16}
may be helpful in understanding the hidden patterns of such structures.
In this section we will walk through the history of the study of voids and
shortly review the most prominent recent works in this field.

In 1978 Gregory and Thompson \cite{Gregory78} and independently of them
Joeveer,  Einasto and  Tago \cite{Joeveer78}, while studying
superclusters, noticed that there are areas in the galaxy field where
the galaxy density is much lower than the average one (Joeveer referred to
them as the ``holes'' in the distribution). Very soon it became obvious that
these holes --- or, as we call them now, cosmic voids --- are very important in the
matter distribution and its evolution. In the 1980s there appeared papers
by renowned cosmologists
(Zeldovich \cite{Zeldovich1982} and Silk \cite{Silk1983}) devoted to the cosmic voids.
They showed that primordial adiabatic cosmological perturbations were the progenitors
of voids, the
same as for the clusters and galaxies, but with the opposite sign. In the same period,
first papers analytically describing the evolution of voids with a certain
profile
were written by Sato \cite{Sato1984} and Maeda \cite{Maeda1983}. The
analytic solutions of the evolution equations for voids predicted already some important
features of the voids, in particular, their expansion during evolution, in
contrast
to the collapse of halos, and the formation of an overdensity shell, the same as we
observe in our numerical solutions. Another important feature, i.e., tendency to
sphericity of the voids, was shown by Fujimoto \cite{Fujimoto1983} and
Icke \cite{Icke1984}.

Unlike the clusters and galaxies, cosmic voids are much harder to track, because they
are underdense and hence unluminous. So, it is not surprising that the $N$-body
cosmological simulation of the large-scale distribution became a powerful tool in
studying the voids, since they have an observational fullness of the void population in
comparison with the real distribution. The first papers describing the voids in a
numerical simulation were written in 1984 \cite{Ryden1984}.

Quite soon, the amount of data on large-scale structure of the Universe
increased quite enough, so that the properties of cosmic voids began being studied
statistically. Already in 1985, Vettolani and colleagues gave statistics on the
volume distribution of 200 real voids \cite{Vettolani1985}. The latest and the
largest galaxy catalogue, on which the void-finding algorithm was imposed, is
SDSS DR12. Mao and colleagues have found more than 10 000 voids there in the
redshift range $0.1<z<0.7$ \cite{Mao2016}.

In the late 1980s the papers discussing the lack of voids in the Ly-alpha forest began
to
emerge \cite{Pierre1988,Ostriker1988}. Due to the faintness of young
galaxies, the discovery of voids at $z>1$ was made only in 1991
\cite{Dobrzycki1991}. The modern searches for early voids in cosmological
simulations were carried out in \cite{Stark2014} and in observable data in
\cite{Ouchi2005}, highlighting, probably, the most distant observable voids at
$z=5.7$ in the spatial distributions of the Lyman-alpha emitters.

Nevertheless, many secrets of the voids' nature were discovered in
distributions obtained from numerical simulations, including the void universal density profile
\cite{Hamaus14}.
The $\Lambda$CDM cosmology simulation with probably the
largest number of voids was done in 2004, where in total 80 000 voids were
found in the simulated distribution \cite{Colberg2005}. Another interesting work
was
written by Gottlober and colleagues \cite{Gottlober2003}, where they studied the
mass distribution in voids, halos in them, the void mass function and its
dependence on the radius using the results of high-resolution $N$-body simulations. With the
development of such simulations, it became possible to track the evolution
of
voids in such simulated universes. The authors of \cite{Arbabi02,
Wojtak2016,SutterElahi2014} studied the changes of profiles and
sizes of the void population with $z$. Modern results on voids from $N$-body
simulations
are given in \cite{SutterElahi2014,Weygaert2016}.

The influence of surroundings on the formation of voids as well as their
hierarchical
structure became clear in 1990s and in early 2000s. Being underdense fragile
structures, the voids can be crushed by the infall of outer regions (i.e., void collapse)
or merge into a larger void.
The first paper describing the void clustering was written in 1994 by  Haque-Copilah
and  Basu \cite{Haque1994}, the interaction of two voids was considered in 1993 by
 Dubinski \cite{Dubinski1993}. In 2003--2004  van de Weygaert and  Sheth
wrote several papers
\cite{ShethWeygaert2003,ShethWeygaert2004}, summing up the studies of the void hierarchy
problem. In the latter paper, the authors managed to formulate the excursion set
formalism of void formation and evolution in the manner like it was previously done for halos
(see also \cite{Clampitt13} and \cite{Jennings13}). They give clear
classification of the types of voids and their density depending on their
neighborhood (void-in-clouds and void-in-voids). They also sum up the
general properties of voids studied at that time.

Starting from the paper \citep{Ghigna1996}, the void statistics is used for
constraining
different cosmological parameters \cite{Dai15}, probing the dark
energy \cite{Biswas10, Bos12,Hamaus15}, testing the modified gravity \cite{Li12,
Li09, Clampitt13, Cai15}. After  Amendola and colleagues computed
the weak gravitational lensing by voids \cite{Amendola1999} it
became possible to find the constraints on cosmological parameters from the
influence of
this lensing on the cosmic microwave background (CMB) \cite{Chantavat2016}
(an example of how voids can be studied
through the gravitational lensing produced by them is given in
\cite{Krause13}). Another important cosmological test, the Alcock-Paczynski
test \cite{APtest}, was predicted and tested in the $N$-body simulation by Lavaux
\cite{Lavaux12} and then confirmed by the observational data by Sutter and team
\cite{SutterAPtest2014}. They used the stacked cosmic voids identified in the SDSS
DR7 and DR10 and measured their ellipticity. The essence of this test is in the
statistical study of the form of large-scale structures. If one uses a
correct cosmology, the relation of sizes along and transverse
the line of sight shows no dependence on $z$. The results confirmed that the
dark energy has no alternatives, but they do not distinguish between the models of
dark energy.

In 2008,  Neyrinck published \cite{Neyrinck2008} the algorithm for finding the
voids
in 3D-galaxy distribution called Zones Bordering On Voidness (ZOBOV). This
algorithm uses the Voronoi tessellation to divide the volume into regions with
some averaged density. The center of each Voronoi cell is one galaxy. After
that, it applies the watershed landform to the obtained 3D density map to find the
edges of voids. ZOBOV algorithm is claimed to be free of parameters and
assumptions about the shapes of expected voids. It became really popular among the
researchers and they now usually use ZOBOV or its modifications (like, for
example, project VIDE \cite{SutterVide}) to find voids in the cosmological simulation
data or real galaxy catalogs.

Despite the far from spherical shape of the walls of the most voids and the influence
of surroundings, the study of the evolution of a spherical isolated void supplies us
with valuable hints in describing the main physical characteristics of the void, sometimes including
astonishingly accurate quantitative guidelines \cite{Weygaert2016}.
The universal density profile of a spherical void has been proposed by
Hamaus, Sutter and  Wandelt \cite{Hamaus14}. The profile has 4 parameters
characterizing the size and
steepness of the void and its overdensity shell. The authors argue that there
is some relation between the form of a void and its size and hence the parameters are not
independent. They show possible dependences that may exist. This work causes a
wide discussion. For example, experts are disputing whether the parameters of
stacked voids profile would be the same \cite{Nadathur16} or different
\cite{Ricciardelli2016} for the simulation and real galaxy distribution.

Finally, we would like to emphasise the role of voids as a probe of
dark energy. Li \cite{Li10} has shown that voids can be
used to rule out observationally or distinguish between coupled scalar field
models of dark energy. Voids are also sensitive to the massive gravity dark energy
models \cite{Spolyar13}. Usually, the shape of voids is studied statistically
to constrain the parameters of dark energy \cite{Biswas10, Bos12, Lavaux10}. Moreover,
with universal density profile, one can improve the Alcock-Paczynski test by
specifying the galaxies' redshift distortion and recovering the shape of voids,
which, again, help to constrain the dark energy parameters \cite{Dai15, Hamaus15}.
Our work is an essential complement to these ones, allowing to follow the evolution
of a single void with all its features.

\section{Spherical voids: the dependence of velocity to density perturbations
ratio on dark energy parameters}

In our previous papers \cite{Tsizh2016,Novosyadlyj2017} we analysed the
formation of
single spherical voids in the three-component medium (matter, radiation and dark
energy) from the early epoch up to the current one. We described each component in the
hydrodynamical approximation. The model of minimally coupled dynamical dark
energy had three free parameters: the energy density in units of the critical one
at
current
epoch\footnote{Hereafter the values of variables at the current epoch are marked with
the upper or lower index 0.} $\Omega_{\text{de}}\equiv8\pi G\varepsilon^0_{\text{de}}/3H_0^2$,
the equation of
state parameter $w_{\text{de}}\equiv p_{\text{de}}/\varepsilon_{\text{de}}$ and the squared effective
sound speed
$c_{\text s}^2\equiv \delta p_{\text{de}}/\delta\varepsilon_{\text{de}}$ in the rest frame of dark
energy. We
assumed that
the last two parameters are constant.
Here and below we use the units in which the speed of light in vacuum $c=1$, all
velocities are in the units of speed of light.

We have shown in \cite{Novosyadlyj2017} how the spherical cosmological
perturbation evolves to form the void with approximately universal density and
velocity profiles. Here, we will show that the ratio of matter velocity and density
perturbations, which can in principle be measured, is sensitive to the
values of main parameters of the dynamical dark energy. This
ratio does not depend on the initial amplitude at the linear stage of evolution.
Since large voids in the spatial distribution of matter, traced by galaxies,
are now at quasi-linear stage of their evolution, the dependence on the initial
amplitude is predictable and can be estimated from the numerical calculations
or $N$-body simulations. On the
other hand, large voids in the Cosmic Web are almost spherical, so
the dependence of this ratio on the profile of initial perturbation and the impact
of
environment can be also removed, at least for the central part of voids. In this
work we only outline the key dependences for the new possible test and leave the
mentioned here important aspects of its implementation for future papers.

\subsection{Assumptions and equations}

Let us consider the evolution of spherical adiabatic cosmological
perturbations
in the Universe filled with the matter, dynamical dark energy and relativistic
component.
The last one is important in the early Universe when we set the initial conditions
following from the observational data on the CMB anisotropy
(see for details \cite{Novosyadlyj2017}).
We present the space-time metric in the region of a perturbation in the Newtonian
gauge as follows:
\begin{equation}
\rd s^2=\re^{\nu(t,r)}\rd t^2-a^2(t)\re^{-\nu(t,r)}\left[\rd r^2+r^2(\rd\theta^2+\sin^2\theta
\rd\varphi^2)\right] \label{ds_sph},
\end{equation}
where $t$ is the cosmological time, $r$ is the radial coordinate comoving to the
unperturbed background, $\theta$ and $\varphi$
are the polar and azimuth angles in the spherical 3D coordinates and $a(t)$ is
the scale factor
of cosmological background. The function $\nu(t,r)$ is the metric perturbation
caused by the local density perturbation of all components. At distances larger than
the
radius of void, $r\gg r_{\text v}$, $\nu(t,r)\rightarrow 0$ and the metric (\ref{ds_sph})
becomes the
Friedmann-Lemaitre-Robertson-Walker (FLRW) one.
The time dependence of the scale factor is determined by the Friedmann equation
\begin{equation}
H^2\equiv\left(\frac{\rd\ln{a}}{\rd t}\right)^2=H_0^2\left[\Omega_{\text{rel}}
a^{-4}+\Omega_{\text{m}} a^{-3}+\Omega_{\text k}
a^{-2}+\Omega_{\text{de}}a^{-3(1+w_{\text{de}})}\right] \label{Fr_eq}.
\end{equation}
In the computations below we set the Hubble constant $H_0$ to be 70~$\text{km}/{\text s}\cdot\text{Mpc}$, this value lies between the ``global'' ($67.8\pm0.9$~$\text{km}/{\text s}\cdot\text{Mpc}$
\cite{Planck2016}) and ``local'' ($73.24\pm0.74$~$\text{km}/{\text s}\cdot\text{Mpc}$
\cite{Riess2016})
measurements. The spatial curvature, as the Planck 2015 results show, is very
close to zero
with $|\Omega_{\text k}|<0.005$, that is why in the most of computations we will put it
equal to zero.

In the spherical perturbed region, the contravariant components of 4-velocity
of any component $u^i_N\equiv \rd x^i/\rd s$ in the metric (\ref{ds_sph}) are
\begin{equation}
u^i_N(t,r) =
\left\{\frac{\re^{-\nu/2}}{\sqrt{1-v_N^2}}\,,\,\,\frac{\re^{\nu/2}v_N}{a\sqrt{1-v_N^2
}}\,,\,\,0,\,\,0\right\}, \label{u^i}
\end{equation}
where the radial component of 3-velocity $v_N$ is defined as the ratio of
proper
space interval to proper time interval at the distance $r$ from the center of the void:
$v_N(t, r )=\re^{-\nu}\rd r/\rd t$ (in units of speed of light).
In the $v_N\ll 1$ approximation, the components of energy-momentum tensor of any
component
$T^i_{j\,(N)}=(\varepsilon_N+p_N)u^i_N(t,r)u_{j\,(N)}-\delta^i_jp_N$ to the
second order of
smallness are as follows:
\begin{align}
T^0_{0\,(N)} &= \varepsilon_N+(\varepsilon_N+p_N)v_N^2\,, \qquad
T^1_{0\,(N)} =(\varepsilon_N+p_N)\frac{\re^{\nu}}{a} v_N\,, \\
T^1_{1\,(N)} &=-p_N-(\varepsilon_N+p_N) v_N^2\,, \qquad  T^2_{2\,(N)} = T^3_{3\,(N)}
=p_N\,.
\label{T^_}
\end{align}
We present the energy density of every component as a sum of the mean (or background)
value
$\bar{\varepsilon}_N(t)$ and the local perturbation:
$\varepsilon_N(t,r)=\bar{\varepsilon}_N(t)[1+\delta_N(t,r)]$, where $\delta_N$
denotes the
relative density perturbation. The pressure of every component in the void
is presented as $p_{N}(t,r)=w_N\bar{\varepsilon}_{N}(t)+\delta p_{N}(t,r)$. For the
scalar field models of dark energy, the non-adiabatic component of pressure perturbation
$\delta p_{N}(t,r)$
is important (see appendix A in \cite{Novosyadlyj2016} and references therein).
Taking this into account, the pressure perturbation of any component in the
generalised form can be presented as
\[\delta p_{N}(t,r)=c^2_{\text{s}\,(N)}\bar{\varepsilon}_{N}\delta_{N}(t,r)-
3\bar{\varepsilon}_{N}aH(1+w_N)\big[c^2_{\text{s}\,(N)}-w_N\big]\int{v_N(t,r)\rd r}.\]
For the relativistic component with $c^2_{\text{s}\,(\text{rel})}=w_{\text{rel}}=1/3$ and the matter one
with
$c^2_{\text{s}\,(\text{m})}=w_{\text{m}}=0$, the non-adiabatic term disappears.

Therefore, in the metric (\ref{ds_sph}) the Einstein equation $G^0_0=\varkappa
T^0_0$ and the conservation
law $T^j_{i\,;j}=0$ give us the system of equations for metric,
density and velocity perturbations of the matter, dark energy and relativistic
components
in the expanding Universe with dynamical dark energy:
\begin{align}
\nu''&+\frac{2}{r}\nu'-3a^2H^2(a\dot\nu+\nu)=
3H^2_0\left[\Omega_{\text m}a^{-1}\delta_{\text{m}}+\Omega_{\text{rel}}a^{-2}\delta_{\text{rel}}+\Omega_{\text{de}}a^{
-(1+3w)}\delta_{\text{de}}\right], \label{nu}\\
\dot{\delta}_{N}&+\frac{3}{a}\left(c_{\text s}^2-w_{N}\right)\delta_{N}+(1+w_{N})\left
[\frac{v'_{N}}{a^2H}+\frac{2v_{N}}{a^2Hr}
-9H\left(c_{\text s}^2-w_{N}\right)\int{v_{N}\rd r}-\frac{3}{2}\dot{\nu}\right] \nonumber \\
&+\left(1+c_{\text s}^2\right)\left[\frac{\delta'_{N}v_{N}}{a^2H}
+\frac{\delta_{N}}{a^2H}\left(v'_{N}+
\frac{2}{r}v_{N}\right)-\frac{3}{2}\delta_{N}\dot{\nu}\right]=0,
\label{delta}\\
\dot{v}_{N}&+\left(1-3c_{\text s}^2\right)\frac{v_{N}}{a}+\frac{c_{\text s}^2\delta'_{N}}{
a^2H(1+w_{N})}+
\left(1+\frac{1+c_{\text s}^2}{1+w_{N}}\delta_{N}\right)\frac{2v_{N}}{a^2H}
\left(v_{N}'+\frac{v_{N}}{r}\right)+\frac{\nu'}{2a^2H}\nonumber\\
&+\frac{1+c_{\text s}^2}{1+w_{N}}\left[\dot{\delta}_{N}v_{N}+\delta_{
N}\dot{v}_{N}+(1-3w_{N})\frac{\delta_{N}}{a}v_{N}+
\frac{\nu'\delta_{N}}{2a^2H}\right]=0. \label{v_m}
\end{align}
We have rewritten the derivatives with respect to the time $\frac{\rd}{\rd t}$ as the
derivatives with respect to the scale factor $\frac{\rd}{\rd t}=aH\frac{\rd}{\rd a}$ and
denoted them as
$(\,\,\dot{ }\,\,)$. The
derivatives with respect to the radial coordinate $\frac{\rd}{\rd r}$ are denoted
here
as $(\,'\,)$.
So, the system
(\ref{nu})--(\ref{v_m}) for 7 unknown functions must be integrated with respect to the scale
factor
$a$ together with the
algebraic equation (\ref{Fr_eq}). To integrate them, the initial conditions must
be set.

\subsection{Initial conditions}

We set the initial conditions in the early epoch when the amplitudes of
perturbations are
small [$\delta_N(a,r)\ll1$, $v_N(a,r)\ll1$ and $\nu(a,r)\ll1$] and the scale of
perturbation is essentially larger than the particle
horizon. For the radiation dominated epoch, we obtain, from the superhorizon asymptotic for the
growth mode
of cosmological perturbations, the relations between initial
amplitudes:
\begin{eqnarray}
\frac{4}{3}\delta_{\text{m}}(a_{\text{init}},r)=\frac{4}{3}\frac{\delta_{\text{de}}(a_{\text{init}},r)}{1+w_{\text{de}}}=\delta_{\text{rel}}(a_{\text{init}},r)=-\nu(a_{\text{init}},r), \label{init_d}\\
v_{\text{m}}(a_{\text{init}},r)=v_{\text{de}}(a_{\text{init}},r)=v_{\text{rel}}(a_{\text{init}},r)=-\frac{\nu'(a_{\text{init}},r)}{
4a_{\text{init}}H(a_{\text{init}})}\,. \label{init_v}
\end{eqnarray}

We set the local initial perturbation as a bell-like hill for the metric function
$\nu(a_{\text{init}},r)=C(1-\kappa^2 r^2)\linebreak\times\exp(-r^2/r_{\text d}^2)$ and an inverted
bell-like
profile for the density, where a free constant $C$ may be linked to the amplitude
of
the power
spectrum of curvature perturbations measured at the last scattering surface of
relic thermal
radiation (see for details \cite{Novosyadlyj2017,Novosyadlyj2016}). We put
here $C=3\cdot10^{-4}$.
We specify other parameters of this profile, which define the initial size
and the slope of wall, as follows: $\kappa^{-1}=50~\text{Mpc}$,
$r_{\text d}=25~\text{Mpc}.$

\begin{figure}[!t]
\includegraphics[width=0.33\textwidth]{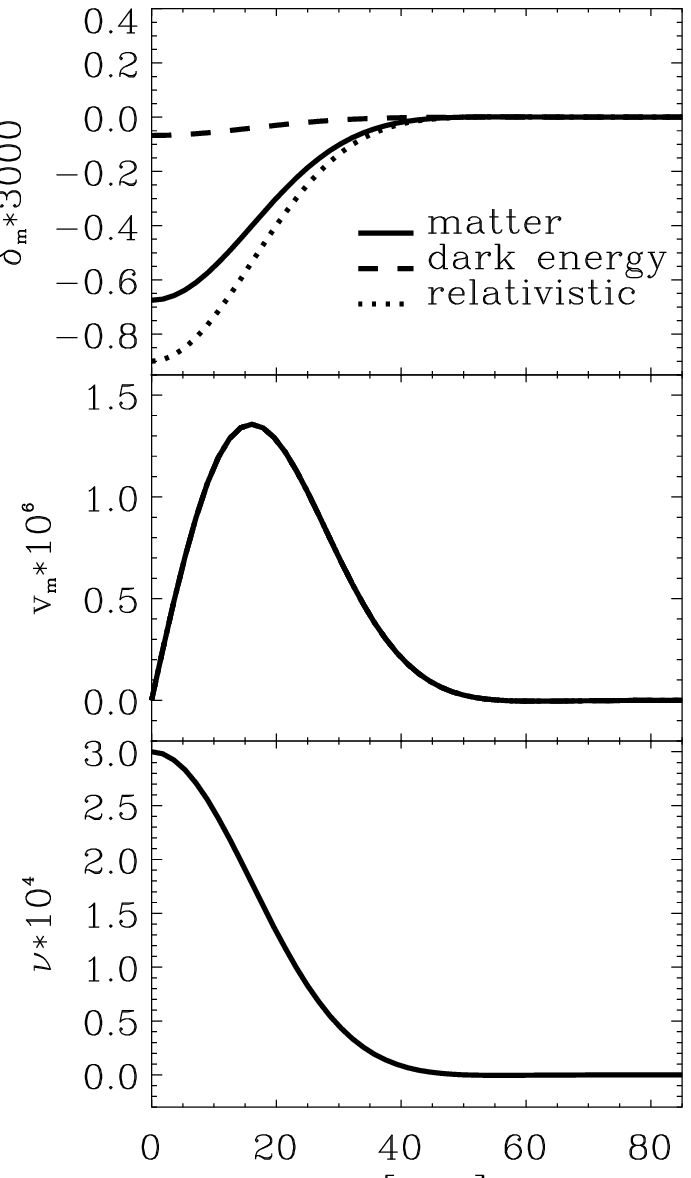}
\includegraphics[width=0.33\textwidth]{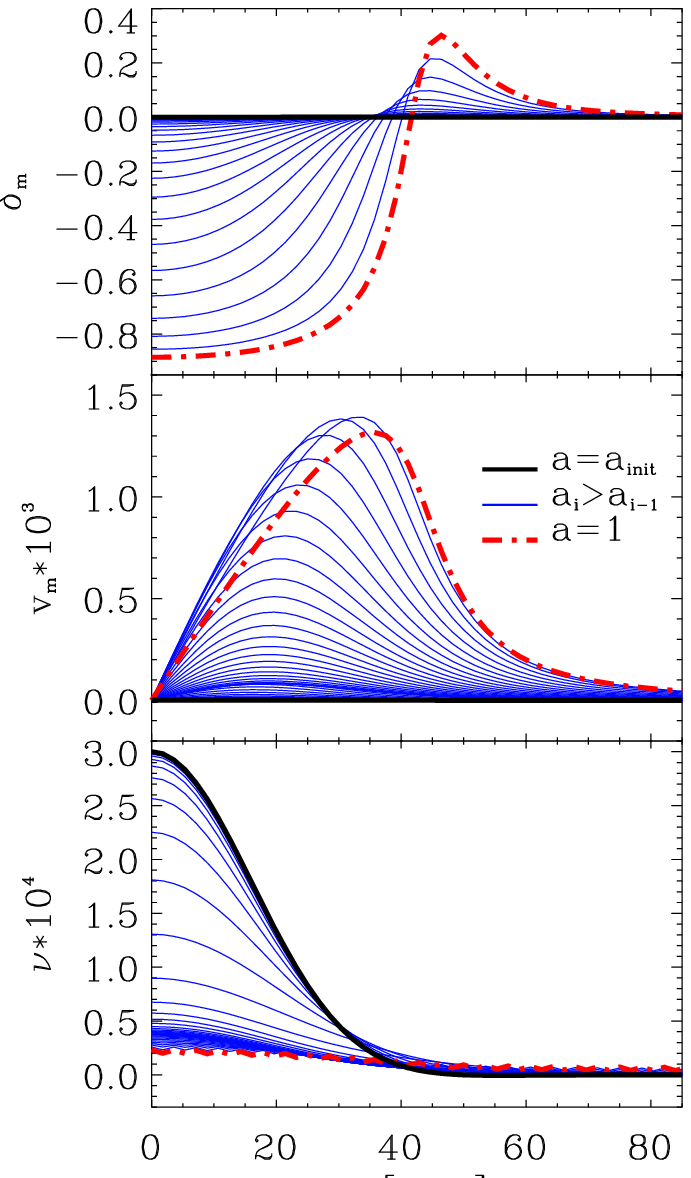}
\includegraphics[width=0.33\textwidth]{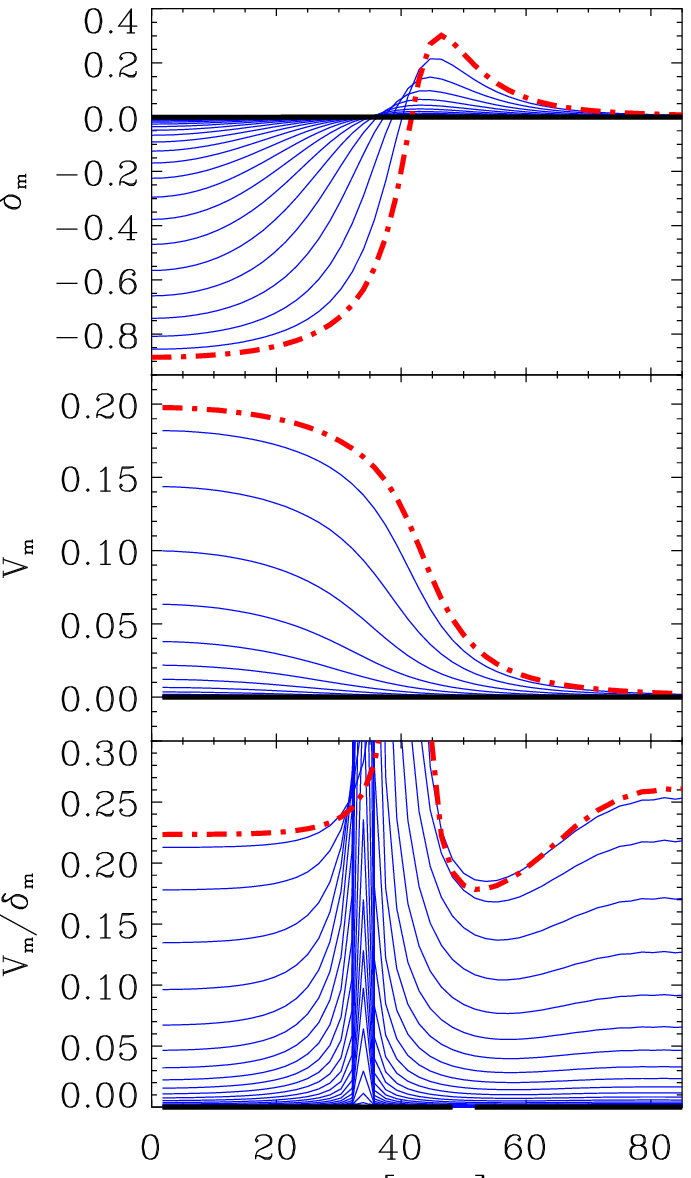}
\caption{(Color online) Formation of the void from the cosmological spherical perturbation.
Left-hand column: the initial profiles of density and velocity perturbations of
matter
and gravitational
potential. Central column: evolution of the density $\delta_{\text{m}}$ and velocity $v_{\text{m}}$
perturbations of matter and gravitational potential $\nu$ from the initial profiles
(thick black lines) to the final one at $a=1$ (thick dashed red lines). Right-hand
column:
evolution of the density perturbation (top), the velocity in units of the Hubble flow
$V_{\text{m}}$ (middle) and the ratio $V_{\text{m}}/|\delta_{\text{m}}|$ (bottom).}
\label{profile_init_evolve}
\end{figure}

The initial profiles of matter density perturbations $\delta_{\text{m}}(a_{\text{init}},r)$,
$\delta_{\text{de}}(a_{\text{init}},r)$,
velocity $v_{\text{m}}(a_{\text{init}},r)\linebreak=v_{\text{de}}(a_{\text{init}},r)=v_{\text{rel}}(a_{\text{init}},r)$ and
gravitational potential
$\nu(a_{\text{init}},r)$ are presented in the left-hand panel of figure~\ref{profile_init_evolve}.

\subsection{Results and discussions}

For numerical integration of the system of equations
(\ref{nu})--(\ref{v_m}) with initial conditions (\ref{init_d})--(\ref{init_v}),
we
have
used the code
\textit{npdes.f}\footnote{\url{http://194.44.198.6/~novos/npdes.tar.gz}; the method of integration described in
\cite{Novosyadlyj2017}.}. The results are presented
in the central panel of figure~\ref{profile_init_evolve}. It illustrates how the
initial spherical matter density perturbation in the form of an inverted bell-like
underdense profile
with initial amplitude $\delta_{\text{m}}(a_{\text{init}},r=0)=-2.25\cdot10^{-4}$ and radius
$r_{\text v}\equiv\kappa^{-1}=50$~Mpc evolves to the spherical void with density contrast
$\delta_{\text{m}}(a=1,r=0)\approx-0.9$ and comoving radius $r_{\delta=0}\approx42$~Mpc.
The matter, which is flowing from the central part of the void outwards, is
collected in its peripheral part, forming an overdense shell and
reducing somewhat the
size of the void in the comoving coordinates.

Maximum of the peculiar velocity of matter is reached at 6/7 of the void final
radius and is
$v_{\text{m}}(1,r_{\text{max\,v}})\linebreak\approx400$~km/s. The value of Hubble flow $V_{\text H}=H_0r$ at this
distance from the
center of the void ($r_{\text{max\,v}}\approx36$~Mpc) equals approximately 2520~km/s.
The ratio $v_{\text{m}}/V_{\text H}$ is $\approx0.15$.
In the middle panel of the right-hand column of figure~\ref{profile_init_evolve}, we
present the
peculiar velocity of matter
in units of the Hubble flow: $V_{\text{m}}\equiv v_{\text{m}}({r})/H_0r$. It has the bell-like form with
maximum $\approx0.20$ in this
special case. Comparing the inverted bell-like form of $\delta_{\text{m}}(r)$ (top
panel)
with the bell-like form of
$V_{\text{m}}(r)$, one suggests that their ratio does not depend on the radius in the central
part of the void. Curves in the bottom panel support this assumption:
$V_{\text{m}}(r)/\delta_{\text{m}}(r)\approx \text{const}$ at any $a$ for
$0\leqslant r\leqslant 0.5r_{\delta=0}$. Let us analyse how this ratio depends on the dark
energy parameters.

\begin{figure}[!t]
\includegraphics[width=0.33\textwidth]{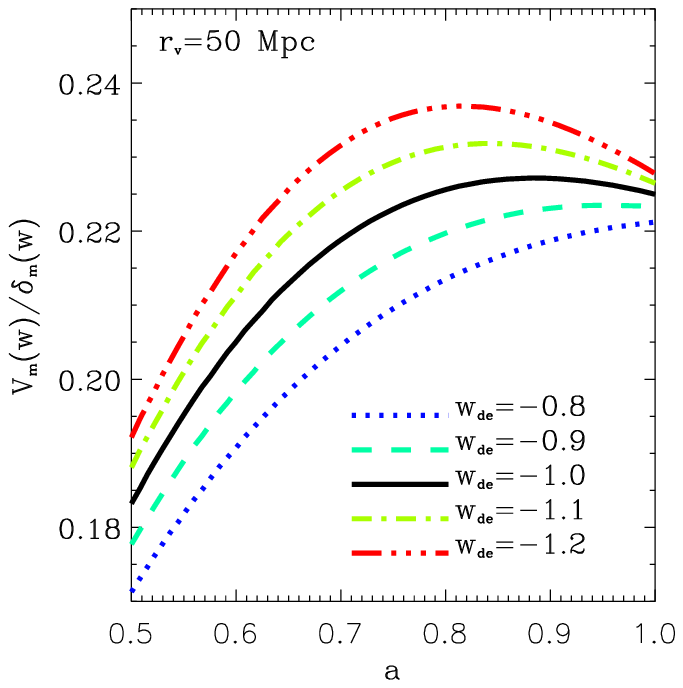}
\includegraphics[width=0.33\textwidth]{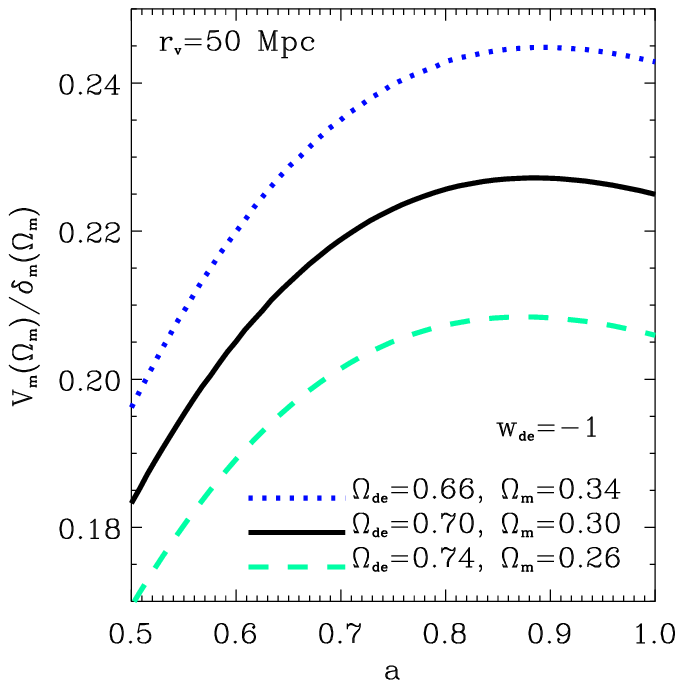}
\includegraphics[width=0.33\textwidth]{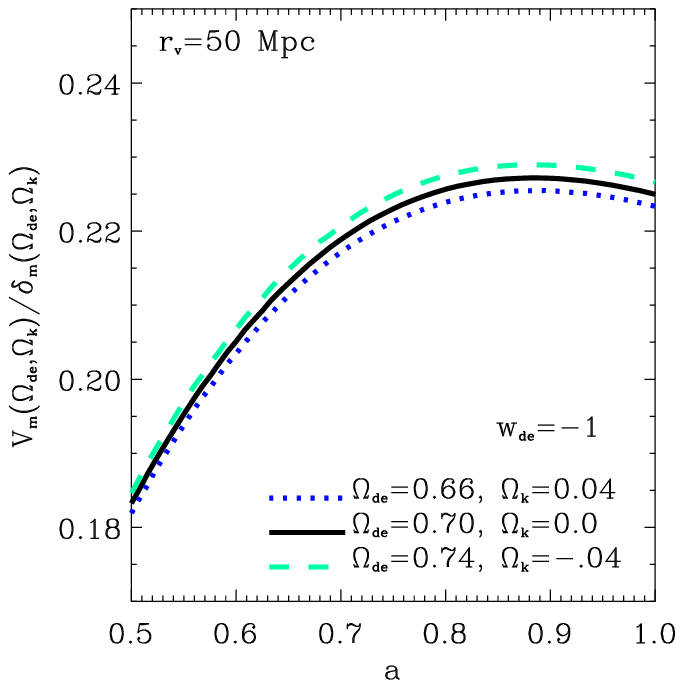}
\caption{(Color online) The ratio $V_{\text{m}}/|\delta_{\text{m}}|$ for dark energy models with different
values
of EoS parameter (left-hand), with different values of density parameter
$\Omega_{\text{de}}$
(central) and fixed zero spatial curvature ($\Omega_{\text k}=0$), with different
values
of density parameter $\Omega_{\text{de}}$ (right-hand) and fixed matter density parameter
($\Omega_{\text{m}}=0.3$). The dark energy perturbations have not been taken into
account.}
\label{statefinder}
\end{figure}
\begin{figure}[!b]
\includegraphics[width=0.33\textwidth]{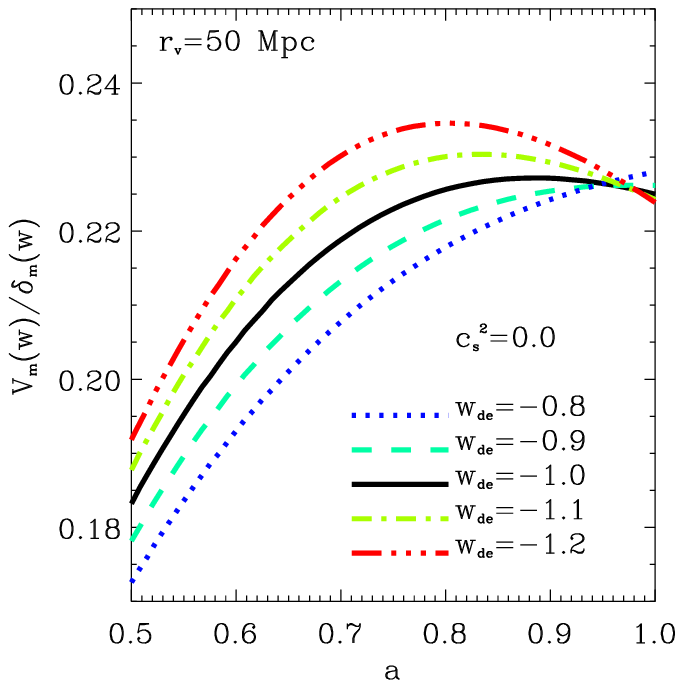}
\includegraphics[width=0.33\textwidth]{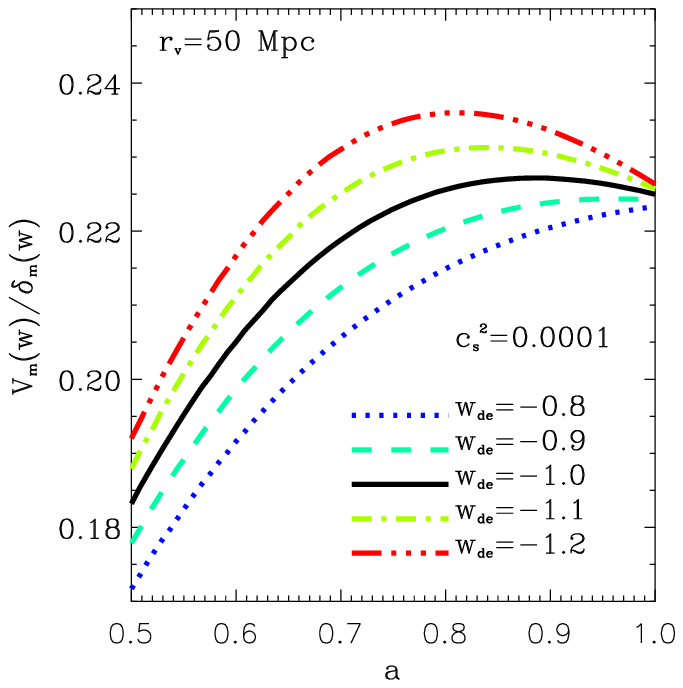}
\includegraphics[width=0.33\textwidth]{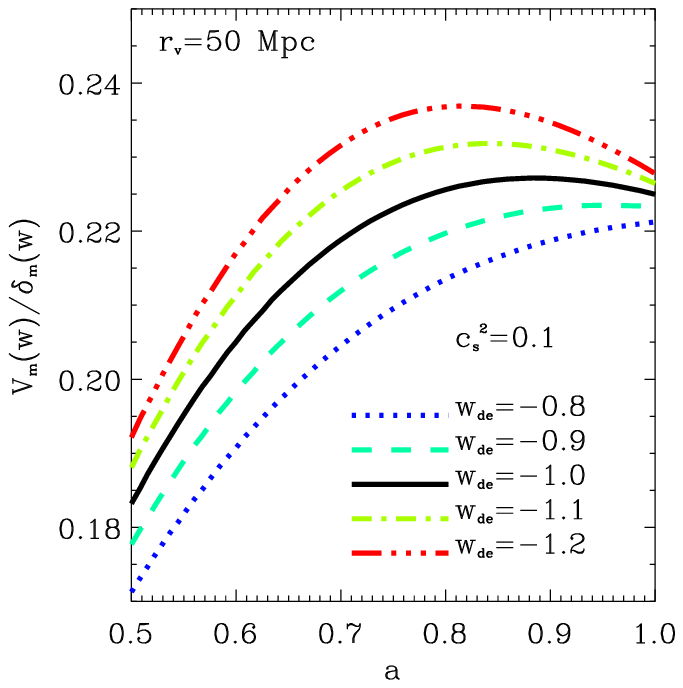}
\caption{(Color online) The ratio $V_{\text{m}}/|\delta_{\text{m}}|$ for dynamical dark energy models with
different values
of effective sound speed. The dark energy perturbations have been taken into
account.}
\label{statefinder2}
\end{figure}

The dark energy impacts the evolution of matter density and velocity
perturbations in two ways.
The first one is realised via the rate of expansion of the Universe: the
equations
(\ref{delta})--(\ref{v_m}) for matter density and velocity perturbations contain the
function
$H(a)$, equation~(\ref{Fr_eq}), which depends on $\Omega_{\text{de}}$ and $w_{\text{de}}$.
The second way consists in the impact of the density perturbations of dark energy on
the
evolution of
matter density and velocity perturbations via the gravitational potential
$\nu$, equation~(\ref{nu}).
The evolution of dark energy density and velocity perturbations, equations~(\ref{delta})--(\ref{v_m}),
depends on $\Omega_{\text{de}}$, $w_{\text{de}}$ and $c_{\text s}^2$. To estimate the level of the impact
of dark energy
on the ratio $V_{\text{m}}(r)/|\delta_{\text{m}}(r)|$ for dark matter we integrate the system of
equations~(\ref{nu})--(\ref{v_m}) in a simplified version ($\nu$, $\delta_{\text{m}}$, $v_{\text{m}}$, $\delta_{\text{rel}}$,
$v_{\text{rel}}$)
and a full version ($\nu$, $\delta_{\text{m}}$, $v_{\text{m}}$, $\delta_{\text{de}}$, $v_{\text{de}}$,
$\delta_{\text{rel}}$,
$v_{\text{rel}}$), without and with the density and velocity perturbations of dark energy.

In figure~\ref{statefinder}, we show the dependences of ratio
$V_{\text{m}}(r)/|\delta_{\text{m}}(r)|$ on scale
factor $a$ when the dark energy perturbations have not been taken into account.
Here and below all dependences of this ratio on $a$ are computed for the point,
which
is at the distance $r=1.8$~Mpc from the central point of the void.
In the left-hand panel we show the dependences of this ratio on scale factor $a$ for the
models of dark
energy with different values of equation of state parameter. One can see that the
lines vary in slope and amplitude for different~$a$. In the central and
right-hand
panels of figure~\ref{statefinder} we show how the ratio $V_{\text{m}}(r)/|\delta_{\text{m}}(r)|$
depends on~$a$ for different values of the dark energy density in the cosmological models
with
fixed zero
spatial curvature ($\Omega_{\text k}=0$) and fixed matter density ($\Omega_{\text{m}}=0.3$), accordingly. In the
case of fixed spatial curvature, the lines correlate with the value of $\Omega_{\text{m}}$
since the equality
$\Omega_{\text{de}}+\Omega_{\text{m}}+\Omega_{\text k}=1$ is always true. The equations~(\ref{nu}) and
(\ref{v_m}) explain
this. When $\Omega_{\text{de}}$ and $\Omega_{\text k}$ are free parameters for fixed
$\Omega_{\text{m}}$ (right-hand panel),
the lines vary less, this illustrates the well-known degeneracy between these
parameters. Since the Planck 2015
results state that $|\Omega_{\text k}|<0.005$ \cite{Planck2016}, the results in the right-hand panel
have only
a theoretical meaning. Therefore, we can state that the observational data on the ratio
$V_{\text{m}}(r)/|\delta_{\text{m}}(r)|$ for the large
voids can be used to constrain the values of density and equation of state
parameters.

Let us integrate now the system of equations~(\ref{nu})--(\ref{v_m}) in the complete
version, when the solutions
for all functions $\nu$, $\delta_{\text{m}}$, $v_{\text{m}}$, $\delta_{\text{de}}$, $v_{\text{de}}$,
$\delta_{\text{rel}}$, $v_{\text{rel}}$ are sought.
The amplitudes of density perturbations of dark energy at the late stages of
evolution depend on the value
of squared effective sound speed $c_{\text s}^2$ of dark energy. They are larger for lower
$c_{\text s}^2$
\cite{Novosyadlyj2017}. The results show that the impact of dark energy
perturbations on the matter
ones is negligibly small for dark energy models with $c_{\text s}^2>0.01$.
It is noticeable only for small values of the effective speed of sound. In figure~\ref{statefinder2} we show
the dependences
of the ratio $V_{\text{m}}(r)/|\delta_{\text{m}}(r)|$ on $a$ for $c_{\text s}^2=0,\,10^{-4},\,0.1$.
One can see that the right-hand plot for $c^2_{\text s}$=0.1 does not practically differ
from the left-hand plot
in figure~\ref{statefinder}, where the perturbations of dark energy have not
been taken
into account at all. The maximal differences of ratios $V_{\text{m}}(r)/|\delta_{\text{m}}(r)|$ for
the corresponding
models in the left-hand panels of
figures~\ref{statefinder} and \ref{statefinder2} are within 2--3~percent. So, the
high accuracy data
of future observations can be used for constraining $c_{\text s}^2$ from below.

Therefore, the observational data on structure of voids and peculiar velocities
of galaxies in them at different redshifts can be used to distinguish the
models of dark energy and determine their parameters.
We understand that these results indicate only the sensitivity of
void characteristics
to the parameters of dark energy models for single spherical voids. Most of the
real voids are non-spherical, surrounded by massive structures, while the galaxies and dark matter are
biased. The impact of the form of the voids,
their surroundings and galactic biasing can be taken into account on the basis
of detailed $N$-body simulations of the formation of a large-scale structure of the
Universe.

\section{Conclusions}

The voids as elements of a large-scale structure of the Universe have been studied
intensively for almost forty years since their discovery. We have learned
many interesting things about their origin, formation, properties and
population, but many problems
connected with them remain unsolved. Huge scientific activity in the world is
connected with the dark energy --- ``the mystery of millennium'', as Thanu
Padmanabhan, renowned
Indian theoretical physicist and cosmologist, said after publication of the results
supporting apparently the viewpoint that something like cosmological constant is inherent for our Universe.
Its
nature remains unknown despite the tremendous efforts of theoretical
physicists,
cosmologists and astrophysicists to shed light into this dark kingdom.
Probably, the
combination of efforts in research of these two kinds of ``darkness'' will help us to
understand the
nature of both. The results presented here nourish this hope.

\newpage
\ukrainianpart

\title{Порожнини в космічній павутині як зонд темної енергії}

\author{Б. Новосядлий, М. Ціж}
\address{Астрономічна обсерваторія, Львівський національний університет імені
Івана Франка, \\
вул. Кирила і Мефодія, 8, 79005 Львів, Україна}

 \makeukrtitle

 \begin{abstract}
 \tolerance=3000%
В роботі досліджено формування великих порожнин з початкових космологічних
адіабатичних збурень метрики простору-часу та густини і швидкості матерії в
моделях Всесвіту з динамічною темною енергією, що зумовлює прискорене
розширення
Всесвіту. Показано, що від'ємні збурення густини з початковим розміром близько
50~Mpc
в супутніх до космологічного фону координатах та амплітудою, що відповідає
спостережуваним
середньоквадратичним флуктуаціям температури мікрохвильового реліктового
випромінювання,
приводять до утворення порожнин з контрастом густини матерії аж до $-$0.9,
максимумом пекулярної
швидкості близько 400~км/c та з радіусом близьким до початкового. Важливою
особливістю досліджених
початкових профілів густини є формування оточуючої оболонки згущення. Показано,
що відношення
пекулярної швидкості в одиницях габблівського потоку до контрасту густини в
центральній частині
порожнини не залежить або слабо залежить від відстані від центра порожнини.
Показано, що таке
співвідношення є чутливим до значень параметрів темної енергії і може бути
використане для їх
знаходження на основі даних про структуру порожнин та пекулярні швидкості
галактик у них.
 \keywords космологічні збурення, темна енергія, великомасштабна структура
Всесвіту, порожнини, космічна павутина

 \end{abstract}

\begin{thebibliography}{50}
%
\bibitem{Einstein1917} Eistein~A., In: Sitzungsberichte der K\"{o}niglich Preu\ss ischen
Akademie der Wissenschaften, Berlin, 1917, 142--152.

\bibitem{Perlmutter1999} Perlmutter~S. \textit{et al.}, Astrophys. J., 1999,
\textbf{517}, 565; \bibdoi{10.1086/307221}.

\bibitem{Riess1998} Riess~A.G. \textit{et al.}, Astron. J., 1998, \textbf{116}, 1009;
\bibdoi{10.1086/300499}.

\bibitem{Schmidt1998}Schmidt B.P. \textit{et al.}, Astrophys. J., 1998, \textbf{507},
46; \bibdoi{10.1086/306308}.

\bibitem{Huterer1999} Huterer D., Turner M.S., Phys. Rev. D, 1999,
\textbf{60}, 081301; \bibdoi{10.1103/PhysRevD.60.081301}.

\bibitem{Amendola2010} Amendola L., Tsujikawa S., Dark Energy: Theory
and Observations, Cambridge University Press, Cambridge, 2010.

\bibitem{Wolschin2010} Lectures on Cosmology: Accelerated Expansion of the Universe, Wolschin G. (Ed.), Springer, Berlin, Heidelberg, 2010;
\bibdoi{10.1007/978-3-642-10598-2}.

\bibitem{Ruiz2010} Dark Energy: Observational and Theoretical Approaches,
Ruiz-Lapuente P. (Ed.), Cambridge University Press, Cambridge, 2010.

\bibitem{Novosyadlyj2013m} Novosyadlyj B., Pelykh V., Shtanov Yu., Zhuk A.,
Dark Energy: Observational Evidence and Theoretical Models,
Akademperiodyka, Kyiv, 2013.

\bibitem{Li09} Li B., Zhao H., Phys. Rev. D, 2009, \textbf{80}, 044027; \bibdoi{10.1103/PhysRevD.80.044027}.

\bibitem{Biswas10} Biswas R., Alizadeh E., Wandelt B.D., Phys. Rev. D, 2010,
\textbf{82}, 023002; \bibdoi{10.1103/PhysRevD.82.023002}.

\bibitem{Bos12} Bos E.G.P., van de Weygaert R., Dolag K., Pettorino V.,
Mon. Not. R. Astron. Soc., 2012, \textbf{426}, 440;\\ \bibdoi{10.1111/j.1365-2966.2012.21478.x}.

\bibitem{Li12} Li B., Zhao G.-B., Koyama K., Mon. Not. R. Astron. Soc., 2012, \textbf{421}, 3481; \bibdoi{10.1111/j.1365-2966.2012.20573.x}.

\bibitem{Clampitt13} Clampitt J., Cai Y.-C., Li B., Mon. Not. R. Astron. Soc., 2013,
\textbf{431}, 749; \bibdoi{10.1093/mnras/stt219}.

\bibitem{Jennings13} Jennings E., Li Y., Hu W., Mon. Not. R. Astron. Soc., 2013, \textbf{434},
2167; \bibdoi{10.1093/mnras/stt1169}.

\bibitem{Cai15} Cai Y.-C., Padilla N., Li B., Mon. Not. R. Astron. Soc., 2015, \textbf{451}, 1036;  \bibdoi{10.1093/mnras/stv777}.

\bibitem{Dai15} Dai D.-C., Mon. Not. R. Astron. Soc., 2015, \textbf{454}, 3590;  \bibdoi{10.1093/mnras/stv2208}.

\bibitem{Hamaus15} Hamaus N., Sutter P.M., Lavaux G., Wandelt B.D.,
J. Cosmol. Astropart. Phys., 2015, \textbf{11}, 036;\\ \bibdoi{10.1088/1475-7516/2015/11/036}.

\bibitem{Tsizh2016} Tsizh M., Novosyadlyj B., Adv. Astron. Space Phys., 2016,
\textbf{6}, 28;	\bibdoi{10.17721/2227-1481.6.28-33}.

\bibitem{Novosyadlyj2017} Novosyadlyj B., Tsizh M., Kulinich Yu., Mon. Not. R. Astron. Soc., 2017,
\textbf{465}, 482; \bibdoi{10.1093/mnras/stw2767}.

\bibitem{Bond96} Bond J.R., Kofman L., Pogosyan D., Nature, 1996, \textbf{380}, 603; \bibdoi{10.1038/380603a0}.

\bibitem{Albert2002} Albert R., Barabasi A.L., Rev. Mod. Phys., 2002, \textbf{74}, 47;
\bibdoi{10.1103/RevModPhys.74.47}.

\bibitem{Holovach06}
  Holovatch Yu., Olemskoi O., von Ferber C., Holovatch T., Mryglod O.,
  Olemskoi I., Palchykov V., J. Phys. Stud., 2006, \textbf{10}, 247.

\bibitem{Holovach16} Holovatch Yu., Kenna R., Thurner S., Preprint \arxiv{1610.01002}, 2016.

\bibitem{Gregory78}
Gregory S.A., Thompson L.A., Astrophys. J., 1978,  \textbf{222}, 784; \bibdoi{10.1086/156198}.

\bibitem{Joeveer78}
Joeveer M., Einasto J., Tago E., Mon. Not. R. Astron. Soc., 1978, \textbf{185}, 357; \bibdoi{10.1093/mnras/185.2.357}.

\bibitem{Zeldovich1982}
Zeldovich Ya.B., Einasto J., Shandarin S.F., Nature, 1982, \textbf{300}, 407; \bibdoi{10.1038/300407a0}.

\bibitem{Silk1983}
Silk J., Szalay A.S., Zel'dovich Ya.B., Sci. Am., 1983, \textbf{249}, No. 4, 56; \bibdoi{10.1038/scientificamerican1083-72}.

\bibitem{Sato1984}
Sato H., In: General Relativity and Gravitation, Bertotti B., de Felice F., Pascolini A. (Eds.), D. Reidel Publishing Company, Dordrecht, 1984, 289--312; \bibdoi{10.1007/978-94-009-6469-3_15}.

\bibitem{Maeda1983}
Maeda K., Sasaki M., Sato H., Prog. Theor. Phys., 1983, \textbf{69}, 89; \bibdoi{10.1143/PTP.69.89}.

\bibitem{Fujimoto1983}
Fujimoto M., Astron. Soc. Jpn., 1983, \textbf{35}, No. 2, 159.

\bibitem{Icke1984}
Icke V., Mon. Not. R. Astron. Soc., 1984, \textbf{206}, 1; \bibdoi{10.1093/mnras/206.1.1P}.

\bibitem{Ryden1984}
Ryden B.S., Turner E.L., Astrophys. J., 1984, \textbf{287}, L59; \bibdoi{10.1086/184398}.

\bibitem{Vettolani1985}
Vettolani G., de Souza R., Marano B., Chincarini G., Astron. Astrophys., 1985, \textbf{144}, No. 2, 506.

\bibitem{Mao2016}
Mao Q. \textit{et al.}, Preprint \arxiv{1602.02771}, 2016.

\bibitem{Pierre1988}
Pierre M., Shaver P.A., Iovino A., Astron. Astrophys., 1988, \textbf{197}, L3.

\bibitem{Ostriker1988}
Ostriker J.P., Bajtlik S., Duncan R.C., Astrophys. J., 1988, \textbf{327}, L35; \bibdoi{10.1086/185135}.

\bibitem{Dobrzycki1991}
Dobrzycki A., Bechtold J., Astrophys. J., 1991, \textbf{377}, L69; \bibdoi{10.1086/186119}.

\bibitem{Stark2014}
Stark C.W., Font-Ribera A., White M., Lee K.-G., Preprint \arxiv{1504.03290}, 2014.

\bibitem{Ouchi2005} Ouchi M. \textit{et al.}, Astrophys. J. Lett., 2005, \textbf{620}, L1; \bibdoi{10.1086/428499}.

\bibitem{Hamaus14}
Hamaus N., Sutter P.M., Wandelt B.D., Phys. Rev. Lett., 2014, \textbf{112}, 251302; \bibdoi{10.1103/PhysRevLett.112.251302}.

\bibitem{Colberg2005}
Colberg J.M., Sheth R.K., Diaferio A., Gao L., Yoshida N., Mon. Not. R. Astron. Soc., 2005, \textbf{360}, 216; \\ \bibdoi{10.1111/j.1365-2966.2005.09064.x}.

\bibitem{Gottlober2003}
Gottl\"ober S., \L okas E. L., Klypin A., Hoffman Y., Mon. Not. R. Astron. Soc., 2003,
\textbf{344}, No. 3, 715;\\ \bibdoi{10.1046/j.1365-8711.2003.06850.x}.

\bibitem{Arbabi02}
Arbabi-Bidgoli S., M\"uller V., Mon. Not. R. Astron. Soc., 2002, \textbf{332}, 205;
\bibdoi{10.1046/j.1365-8711.2002.05296.x}.

\bibitem{Wojtak2016}
Wojtak R., Powell D., Abel T., Mon. Not. R. Astron. Soc., 2016, \textbf{458}, 4431;
\bibdoi{10.1093/mnras/stw615}.

\bibitem{SutterElahi2014}
Sutter P.M., Elahi P., Falck B., Mon. Not. R. Astron. Soc., 2014, \textbf{445}, No. 2, 1235; \bibdoi{10.1093/mnras/stu1845}.

\bibitem{Weygaert2016}
Van de Weygaert R., Proc. Int. Astron. Union, 2016,
\textbf{11}, No. S308, 493;
\bibdoi{10.1017/S1743921316010504}.

\bibitem{Haque1994}
Haque-Copilah S., Basu D., Publ. Astron. Soc. Pac., 1994, \textbf{106}, No. 695, 67; \bibdoi{10.1086/133344}.

\bibitem{Dubinski1993}
Dubinski J., da Costa L.N., Goldwirth D.S., Lecar M., Piran T., Astrophys. J., 1993, \textbf{410}, No. 2, 458;\\ \bibdoi{10.1086/172762}.

\bibitem{ShethWeygaert2003}
Van de Weygaert R., Sheth R.K., Preprint \arxiv{astro-ph/0310914}, 2003.

\bibitem{ShethWeygaert2004}
Sheth R.K., van de Weygaert R., Mon. Not. R. Astron. Soc., 2004, \textbf{350}, No. 2, 517;\\ \bibdoi{10.1111/j.1365-2966.2004.07661.x}.

\bibitem{Ghigna1996}
Ghigna S., Bonometto S., Retzlaff J., Gottloeber S., Murante G., Astrophys. J., 1996, \textbf{469}, 40; \bibdoi{10.1086/177755}.

\bibitem{Amendola1999}
Amendola L., Frieman J., Waga I., Mon. Not. R. Astron. Soc., 1999, \textbf{309}, No. 2, 465;\\  \bibdoi{10.1046/j.1365-8711.1999.02841.x}.

\bibitem{Chantavat2016}
Chantavat T., Sawangwit U., Sutter P.M., Wandelt B.D., Phys. Rev. D, 2016, \textbf{93}, 043523;\\ \bibdoi{10.1103/PhysRevD.93.043523}.

\bibitem{Krause13}
Krause E., Chang T.-C., Dor\'{e} O., Umetsu K., Astrophys. J. Lett., 2013, \textbf{762}, L20; \bibdoi{10.1088/2041-8205/762/2/L20}.

\bibitem{APtest} Alcock C., Paczy\'nski B., Nature, 1979, \textbf{281}, 358;
\bibdoi{10.1038/281358a0}.

\bibitem{Lavaux12} Lavaux G., Wandelt B.D., Astrophys. J., 2012, \textbf{754},
109; \bibdoi{10.1088/0004-637X/754/2/109}.

\bibitem{SutterAPtest2014}
Sutter P.M., Pisani A., Wandelt B.D.,  Weinberg D.H., Mon. Not. R. Astron. Soc., 2014, \textbf{443}, 2983;\\ \bibdoi{10.1093/mnras/stu1392}.

\bibitem{Neyrinck2008}
Neyrinck M., Mon. Not. R. Astron. Soc., 2008, \textbf{386}, 2101; \bibdoi{10.1111/j.1365-2966.2008.13180.x}.

\bibitem{SutterVide}
Sutter P.M., Lavaux G., Hamaus N., Pisani A., Wandelt B.D., Warren M., Villaescusa-Navarro F., Zivick P., Mao Q., Thompson B.B., Astron. Comput., 2015, \textbf{9}, 1; \bibdoi{10.1016/j.ascom.2014.10.002}.

\bibitem{Nadathur16}	
Nadathur S.,  Hotchkiss S., Diego J.M., Iliev I.T., Gottl\"{o}ber S., Watson W.A., Yepes G., Proc. Int. Astron. Union, 2016, \textbf{11}, No. S308, 342; \bibdoi{10.1017/S1743921316010541}.

\bibitem{Ricciardelli2016}
Ricciardelli E., Quilis V., Varela J., Proc. Int. Astron. Union, 2016, \textbf{11}, No. S308, 551;\\ \bibdoi{10.1017/S1743921316010565}.

\bibitem{Li10} Li B., Mon. Not. R. Astron. Soc., 2011, \textbf{411}, 2615;
\bibdoi{10.1111/j.1365-2966.2010.17867.x}.

\bibitem{Spolyar13} Spolyar D., Sahl\'en M., Silk J., Phys. Rev. Lett., 2013,
\textbf{111}, 241103; \bibdoi{10.1103/PhysRevLett.111.241103}.

\bibitem{Lavaux10} Lavaux G., Wandelt B.D., Mon. Not. R. Astron. Soc., 2010, \textbf{403}, 1392;
\bibdoi{10.1111/j.1365-2966.2010.16197.x}.

\bibitem{Planck2016} Planck Collaboration, Astron. Astrophys., 2016,
\textbf{594}, A13;  \bibdoi{10.1051/0004-6361/201525830}.

\bibitem{Riess2016} Riess A. \textit{et al.}, Astrophys.
J., 2016, \textbf{826}, 56; \bibdoi{10.3847/0004-637X/826/1/56}.

\bibitem{Novosyadlyj2016} Novosyadlyj B., Tsizh M., Kulinich Yu., Gen. Relativ. Gravitation, 2016,
\textbf{48}, 30; \bibdoi{10.1007/s10714-016-2031-8}.

\end{thebibliography}
\end{document}